\documentclass[a4paper,11pt]{article}
\usepackage{pos-arxiv}

\newcommand{\be}{\begin{equation}}
\newcommand{\ee}{\end{equation}}

\title{Quantum fields, strings, and physical mathematics}

\author{Piotr Su{\l}kowski}

\affiliation{Faculty of Physics, University of Warsaw, \\
ul. Pasteura 5, 02-093 Warsaw, Poland}
  
 \affiliation{Walter Burke Institute for Theoretical Physics, California Institute of Technology, \\
 Pasadena, CA 91125, USA}

\emailAdd{psulkows@fuw.edu.pl}

\abstract{I briefly review several important formal theory developments in quantum field theory and string theory that were reported at ICHEP conferences in past decades, and explain how they underlie a new research area referred to as physical or quantum mathematics. To illustrate these ideas in some specific context, I discuss certain aspects of topological string theory and a recently discovered knots-quivers correspondence.
}

\FullConference{%
  40th International Conference on High Energy Physics -- ICHEP2020\\
  July 28 -- August 6, 2020\\
  Prague, Czech Republic (virtual meeting)
}

\begin{document}
\maketitle


\section{Introduction}

This year's International Conference on High Energy Physics (ICHEP) is exceptional for several reasons. It is the 40th ICHEP conference, taking place 70 years after the first meeting in this series in Rochester in 1950. These anniversaries provide a great motivation to summarize some highlights in high energy physics from the last 70 years and to contemplate how they underlie the current developments. In this note I follow such a plan in the context of ``formal theory developments'', discussed in a separate session at ICHEP 2020. Of course, in limited space and time, it is impossible to present all important contributions in this area. Therefore I focus on results related to the research area referred to as \emph{physical mathematics} or \emph{quantum mathematics}, which has been developed recently very actively. While the scope and the boundaries of this field are not so clearly defined, it is rooted in formal developments in quantum field theory and string theory, in particular those that were reported in the past ICHEP meetings. To illustrate these ideas in a specific context, I also discuss some aspects of topological string theory, as well as recently discovered knots-quivers correspondence.

The plan of this note is as follows. In section \ref{sec-ichep} I briefly review several important ``formal theory developments'' in quantum field and string theory discussed in past ICHEP meetings, which underlie various current research areas. In section \ref{sec-phys-math} I summarize what physical or quantum mathematics is supposed to be. In section \ref{sec-threads} I discuss a crucial role of these ideas in topological string theory. In section \ref{sec-knots} I explain how these ideas get unified in the \emph{knots-quivers correspondence}, a current research direction that can be thought of as one representative of physical or quantum mathematics.


\section{Formal theory developments and 40 ICHEP meetings}  \label{sec-ichep}

Since 1950 most important developments in high energy physics, including those related to theoretical and mathematical structure of quantum field theory and string theory, have been reported during ICHEP conferences. To illustrate at least some of these developments let us recall a few important ICHEP summary talks, which presented ideas that underlie later development of the research area referred to as physical or quantum mathematics. 

One such important theoretical idea is that of large $N$ expansion in $SU(N)$ gauge theories. Intriguing properties of large $N$ expansion were discovered by 't Hooft, who hoped that this approach would lead to a rigorous formulation of quantum gauge theories. In his ICHEP talk in 1982, entitled \emph{``Theoretical perspectives''} \cite{tHooft:1982hlm}, he presented his vision of the future of high energy physics, which in \emph{``Stage 5, still further in the future, may finally give us a convergent field theory. This is a field theory that is mathematically as rigorous as presently known constructive field theories in 2 or 3 space-time dimensions''}. He then claimed that his \emph{``own work now points towards a conjecture: $SU(N\to\infty)$ gauge theories may be mathematically well defined''}. While rigorous formulation of gauge theories is still beyond our reach, the idea of considering the large $N$ limit proved very fruitful in various contexts. For example, it led to powerful methods of solving random matrix models, which can be thought of as 0-dimensional gauge theories. In large $N$ limit one can introduce a topological expansion of matrix model amplitudes, ribbon diagrams, loop equations and their solution in terms of the topological recursion (recently further generalized to Airy structures), etc. Matrix models and their large $N$ limit are also intimately related to gauge theories with extended supersymmetry in various dimensions (e.g. in Dijkgraaf-Vafa theory), Chern-Simons theory, and topological string theory. Furthermore, the vast field of AdS/CFT correspondence also relies on properties of the large $N$ limit. It is hard to imagine modern theoretical high energy physics without developments that rely on the analysis of large $N$ limit in gauge and string theories.

Another important idea bringing together physics and mathematics was presented by D. Gross in his ICHEP 1988 talk \emph{``Superstrings and Unification''} \cite{Gross:1988jb}. Apart from developments in string theory, in his talk he also explained that in \emph{``very recent work, Witten has opened up the subject in a totally new direction. It turns out that two-dimensional conformal field theory is connected with three-dimensional general relativity! If you take as a three-dimensional action the Chern-Simons term: $S=k\int d^3x \epsilon^{ijk} \textrm{Tr} (A_i \partial_k A_j + \frac23 A_i A_j A_k)$, you obtain a gauge invariant and generally covariant action. It is generally covariant without containing the metric explicitly thus it is a \emph{topological field theory} in which the gauge invariant observables cannot depend on the metric and thus must be topological invariants. The gauge field $A_i$ is a matrix in the Lie algebra of some group, say $SU(N)$. Now, if you construct the Wilson loop around a curve, or knot in three dimensions you get a topological knot invariant, i.e. a number that depends only on the topology of the knot and on $N$ and $k$. Witten demonstrated that these invariants coincide with [...] 
the Jones polynomials''}. Since then relations between gauge theory and knot theory have been generalized to string theory and their analysis has grown into an important research direction, as we also review in what follows. 

Furthermore, in ICHEP 1996 in Warsaw, in \emph{``Recent developments in non-perturbative quantum field theory''} talk \cite{Ferrara:1996fc}, S. Ferrara did \emph{``summarize some of the main basic results of the years 94-96, in the context of string theory and its non-perturbative regime''}. Among others, he mentioned that \emph{``The Seiberg-Witten solution of rigid $N=2$ theory generalizes to heterotic-type II duality''}, \emph{``Witten proved the equivalence of different string theories in higher dimensions and the duality of type IIA at strong coupling with 11D supergravity at large radius (M-theory on $M_{10}\times S^1)$''}, and that \emph{``M-theory and strings may undergo a further unification in twelve dimensions (F-theory)''}. Note that all these developments rely in a crucial way on extended supersymmetry.

Formal aspects of high energy physics were also summarized in two talks in ICHEP 2000. M. Dine presented \emph{``Recent Progress in Field Theory''} \cite{Dine:2000zy}, admitting that \emph{``Yet field theory also has limitations. We probably need to go beyond quantum field theory if we are to understand the problems of black holes, the cosmological constant problem, the principles which determine the ground states of M-theory and what selects among them''}. P. Townsend summarized developments in his talk \emph{``Superstring and supermembrane theory''} \cite{Townsend} that he finished with the wish \emph{``May the next 20 years be as fruitful!''}, which shows the enthusiasm concerning these topics at that time.

The last historical talk that we mention is by A. Sen in ICEHP 2010, entitled \emph{``String Theory: Basic Facts and Recent Developments''} \cite{Sen:2010zzg}. Among others, he listed \emph{``some examples of recent developments in quantum field theories. 1. We now have exact results for anomalous dimensions of operators in $\mathcal{N}=4$ super Yang-Mills theories in the planar limit [...]. 2. In another line of development, many exact results for on-shell S-matrix elements in $\mathcal{N}=4$ supersymmetric Yang-Mills theories have been derived using on-shell methods [...]. 3. In a remarkable series of papers Gaiotto and his collaborators developed tools to construct and study a whole new class of $\mathcal{N}=2$ superconformal field theories in four dimensions -- some without even a Lagrangian description [...] 4. Possible finiteness of $\mathcal{N}=8$ supergravity has been an active area of research [...]''}.

From several talks mentioned above we can immediately identify important formal developments in quantum field theory, superstring theory, and M-theory, such as understanding of dualities and non-perturbative effects, topological invariance, large $N$ limit, etc., which underlie the field of physical mathematics. In particular, extended supersymmetry is an important ingredient in most of these developments -- on one hand it enables to keep non-perturbative effects under control, and on the other hand it may be used to impose topological invariance in field or string theories.


\section{Physical and quantum mathematics}   \label{sec-phys-math}

The terms physical mathematics and quantum mathematics have been around for some time. They are not rigorously defined and their meaning may depend a bit on a context and a user; quite often -- though not always -- they are used interchangeably. In a broad sense, they refer to considerations motivated by physics, conducted in the formalism of quantum field theory or string theory, and relying on physical arguments, which lead to (unexpected, non-trivial, and often conjectural) statements in mathematics. Quite often such statements arise from the analysis of some physical duality relating two systems, which have independent mathematical descriptions. A duality of these systems leads then to new relations between various mathematical objects that originate from completely different theories. It is instructive to recall how the idea of physical mathematics was presented on several other occasions. 

For example, during his summary talk at Strings conference in 2014 \cite{Moore}, G. Moore explained that \emph{``after 40-odd years of a flowering of intellectual endeavor a new field has emerged with its own distinctive character, its own aims and values, its own standards of proof. I like to refer to the subject as Physical Mathematics. [...] The use of the term “Physical Mathematics” in contrast to the more traditional \emph{Mathematical Physics} by myself and others is not meant to detract from the venerable subject of Mathematical Physics but rather to delineate a smaller subfield characterized by questions and goals that are often motivated, on the physics side, by quantum gravity, string theory, and supersymmetry, (and more recently by the notion of topological phases in condensed matter physics), and, on the mathematics side, often involve deep relations to infinite-dimensional Lie algebras (and groups), topology, geometry, and even analytic number theory, in addition to the more traditional relations of physics to algebra, group theory, and analysis. 
[...] one of the guiding principles is the goal of understanding the ultimate foundations of physics. [...] 
If a physical insight leads to a significant new result in mathematics, that is considered a success. It is a success just as profound and notable as an experimental confirmation from a laboratory of a theoretical prediction of a peak or trough. For example, the discovery of a new and powerful invariant of four-dimensional manifolds is a vindication just as satisfying as the discovery of a new particle.''}

In a similar vein, M. Atiyah, R. Dijkgraaf and N. Hitchin in \cite{ADH} \emph{``review the remarkably fruitful interactions between mathematics and quantum physics in the past decades, pointing out some general trends and highlighting several examples, such as the counting of curves in algebraic geometry, invariants of knots and four-dimensional topology''}.

Finally, in \cite{Zaslow:2005jw} E. Zaslow characterizes such an activity and developments as \emph{``Physmatics''}.


\section{A playground -- topological string theory} \label{sec-threads}

Above we reviewed several highlights presented during past ICHEP conferences. In this section we briefly describe how these phenomena, originally discovered independently, play a crucial role in topological string theory, which itself is an important playground of physical mathematics.


\subsection{Topological string theory, large $N$, and geometric transitions}

Topological string theory is a simplified version of string theory, whose playground is a Calabi-Yau threefold (of 6 real dimensions). Its definition involves a two-dimensional $\mathcal{N}=2$ supersymmetric sigma-model, whose appropriate twists impose topological invariance and result in A-model or B-model version of the theory \cite{Hori:2003ic}. In this sense extended supersymmetry, as an important ingredient of physical mathematics, also plays a fundamental role in topological strings. As in ordinary string theory, one can consider closed and open version of topological strings. For a Calabi-Yau threefold $M$ with K{\"a}hler parameters $Q_i=e^{t_i}$ characterizing its 2-cycles, and $Q=\{Q_i\}$, closed topological string partition function has the following expansion in the string coupling $g_s$
\be
Z^{\textrm{closed}} = \exp \Big(\sum_{g=0}^{\infty} g_s^{2g-2} F_g(Q)  \Big),   \label{Zclosed}
\ee
where $F_g(Q)$ are referred to as genus $g$ free energies and, mathematically, they encode Gromov-Witten invariants of $M$. In particular, the leading contribution $F_0(Q)$ encodes classical triple intersection numbers of $M$. On the other hand, open topological string partition function takes  form
\be
Z^{\textrm{open}} = \exp \Big(\sum_{n=0}^{\infty} g_s^{n-1} S_n(x,Q)  \Big),   \label{Zopen}
\ee
where $S_n(x,Q)$ depend both on closed K{\"a}hler parameters $Q$ and open parameters $x$ that can be thought of as characterizing (topological) branes. 

In topological strings, large $N$ limit arises in the context of an equivalence of open and closed versions of the A-model theory on two appropriately chosen Calabi-Yau manifolds. In appropriate setup, open A-model theory turns out to have an effective description in terms of $SU(N)$ Chern-Simons theory, where $N$ characterizes boundary conditions, which can be interpreted as being imposed by $N$ branes wrapping a lagrangian cycle \cite{Witten:1992fb}. It is in this sense that we can consider large $N$ limit. Note that various observables in Chern-Simons theory, in the large $N$ limit, indeed have topological expansion analogous to (\ref{Zclosed}) or (\ref{Zopen}); in particular, contributions $F_g(Q)$ are encoded in ribbon diagrams of genus $g$. Furthermore, it turns out that in the large $N$ limit the open A-model theory on one manifold can be interpreted as closed version of the theory on another manifold, with the above mentioned lagrangian cycle replaced by a 2-cycle, whose K{\"a}hler parameter is given by the 't Hooft coupling $t=g_sN$. The process of replacing one of these threefolds by another one is also referred to as a geometric transition, and it provides a nice example of an open-closed duality.

The simplest and prototype example of a geometric transition arises when open and closed topological strings are considered for two versions of a conifold geometry: deformed and resolved one, which are two different deformations of a singular conifold (defined by a complex equation $z_1^2+\ldots +z_4^2=0$ in $\mathbb{C}^4$ parametrized by $z_i$). The deformed conifold is simply a cotangent bundle $T^*S^3$, where we can consider topological branes wrapped on the lagrangian $S^3$. The resolved conifold is a bundle $\mathcal{O}(-1)\oplus\mathcal{O}(-1)\to \mathbb{P}^1$, with a 2-cycle $\mathbb{P}^1$. Open opological strings on the deformed conifold with $N$ branes on $S^3$ are described by $SU(N)$ Chern-Simons theory on $S^3$; in particular, in this setup, topological string partition function is equal to Chern-Simons partition function on $S^3$. Upon the geometric transition, deformed conifold with $N$ branes on $S^3$ is transformed into resolved conifold with $\mathbb{P}^1$ of size $t=g_sN$, and appropriate observables in these two theories are equal \cite{Gopakumar:1998ki}. 

Note that the geometric transition provides an interesting relation, in the spirit of physical mathematics, between two mathematical fields: low-dimensional topology (which provides invariants, such as Witten-Reshetikhin-Turaev invariants, that characterize 3-manifolds wrapped by $N$ branes before the transition), and Gromov-Witten theory (which provides a mathematical description of closed topological string theory after the transition).


\subsection{From Chern-Simons theory to knot homologies}

As mentioned in section \ref{sec-ichep}, Chern-Simons theory with Wilson loop observables computes polynomial knot invariants, such as colored HOMFLY-PT polynomials $P_R(a,q)$ or colored Jones polynomials $J_R(q)=P_R(a=q^2,q)$. From gauge theory perspective, the parameter $a=q^N$ encodes the rank of the $SU(N)$ Chern-Simons gauge group, while $q=\exp\frac{2\pi i}{k+N}$ depends on $N$ and the Chern-Simons level $k$. The color $R$ denotes a representation of $SU(N)$, identified with a Young diagram. For the fundamental representation $R=\square$, colored polynomials reduce to ordinary HOMFLY-PT or Jones polynomials. In what follows we focus on symmetric representations $R=S^r$, labeled by Young diagrams made of one row, and denote $P_{S^r}(a,q)\equiv P_r(a,q)$.

It turns out that the relation between knot invariants and gauge theory observables can be also lifted to topological string theory. To this end, in the open A-model theory on $T^*S^3$, we have to include an additional brane, which intersects $S^3$ along a knot of interest. In this case open topological string amplitudes reduce to Chern-Simons amplitudes, and indeed reproduce knot invariants of this knot. A configuration with such an additional brane can also undergo a geometric transition, which produces a resolved conifold with an extra brane that also captures properties of the knot. In this case, open string partition function, which after the transition encodes open Gromov-Witten invariants according to (\ref{Zopen}), takes form of a generating function of colored HOMFLY-PT polynomials which are computed by Chern-Simons theory (before the geometric transition) 
\be
Z^{\textrm{open}} = \sum_{r=0}^{\infty} P_r(a,q)x^r,   \label{Z-homfly}
\ee
where only symmetric representations arise, $x$ is an open K{\"a}hler parameter, $q=e^{g_s}$, and $a=e^{g_sN}$ is identified with the closed K{\"a}hler parameter of the resolved conifold $Q$ in (\ref{Zopen}).

Furthermore, polynomial knot invariants, such as Jones or HOMFLY-PT polynomials, turn out to have much deeper meaning and structure. Namely, they arise as Euler characteristics of various homological theories associated to knots. A prominent example of such homological construction is Khovanov homology, which categorifies Jones polynomial (i.e. Jones polynomial arises as an Euler character of Khovanov homology). There are many other knot homologies associated to other polynomial knot invariants; some of them are quite abstract and difficult to construct in practice, while in some other cases we can at most predict certain expected properties of such homologies (without providing their explicit construction), as is the case for HOMFLY-PT homology. 

Amusingly, homological knot invariants can be also related to high energy physics and various constructions in string theory, in particular to the brane system mentioned above \cite{Gukov:2004hz}. In this case, it is postulated that homological spaces should be identified with spaces of BPS arising in this brane system, as we review in a little more detail below. While explicit identification of BPS or homological spaces from this viewpoint is difficult and often still impossible, this physical approach enables for example computation of superpolynomials, which are Poincare (rather than Euler) characteristics of knot homologies and do capture certain homological information. Superpolynomials for various knots can be computed using various approaches: refined topological vertex, refined Chern-Simons theory, relation to DAHA, or expected structural properties of knot homologies. There are also other relations between gauge theory and string theory and knots homologies, for example see \cite{WItten:2011pz}.


\subsection{Topological string theory and BPS counting}

As mentioned above, an important aspect of topological string theory is its relation to knot homologies, which can be identified with spaces of BPS states. Let us recall how these spaces of BPS states arise. To identify them, one should embed a topological string setup in the full 10-dimensional superstring theory or in M-theory. Focusing on the latter system, we consider 11-dimensional space, whose 6 dimensions are identified with a Calabi-Yau threefold considered above (e.g. deformed or resolved conifold), 4 dimensions form ``spacetime'' $\mathbb{R}^4$, and the 11th dimension is $S^1$. In addition, we include in this setup topological branes mentioned above that are identified as parts of M5-branes, which extend along 3 dimensions inside Calabi-Yau, span $\mathbb{R}^2$ subspace of $\mathbb{R}^4$, and wrap the M-theory circle $S^1$. In particular, a lagrangian brane that engineers a knot in a conifold is identified as a 3-dimensional part of the full M5-brane. 

BPS states in such M-theory systems are represented by M2-branes, and corresponding partition functions encode multiplicities of such BPS states. More precisely, closed BPS states are represented by M2-branes that wrap 2-cycles inside Calabi-Yau \cite{Gopakumar:1998ii}. On the other hand, open BPS states arise in systems with M5-branes; such open BPS states are represented by open M2-branes  whose boundaries are attached to M5-branes \cite{Ooguri:1999bv}. Furthermore, it is expected that such systems are represented by effective supersymmetric quantum gauge theories in directions normal to Calabi-Yau space. For systems with M5-branes, such an effective theory may also live in 3 dimensions of M5-brane normal to Calabi-Yau space, i.e. $\mathbb{R}^2$ spacetime directions and extra $S^1$ of the M5-brane. 

To be more specific, this BPS interpretation constrains the form of closed topological string partition function (\ref{Zclosed}) as follows
\be
Z^{\textrm{closed}}(Q) = \prod_{\beta\in H_2(M)}  \prod_j\prod_{l=1}^{\infty} (1 - Q^{\beta} q^{l+j})^{l N_{\beta,j}},    \label{Z-closed-GV}
\ee
where $N_{\beta,j}$ are conjecturally integer Gopakumar-Vafa invariants that count BPS states of closed M2-branes and $q=e^{g_s}$. For a fixed $\beta$ and $j$, the contribution from the product over $l$ is a generalization of the MacMahon function $M(q)=\prod_{l=1}^{\infty}(1-q^l)^{-l}$ that counts plane partitions. There is an analogous expression for open topological string partition function (\ref{Zopen}). For simplicity, assuming that it depends on a single open parameter $x$, it reads
\be
Z^{\textrm{open}}(Q,x) = \prod_{j,\beta,k} \prod_{l=1}^{\infty} (1 - x^{k} Q^{\beta} q^{l+j-1/2}  )^{N_{\beta,k,j}},   \label{ZUV-prod}
\ee
where $N_{\beta,k,j}$ count open BPS states and are referred to as Ooguri-Vafa or (in case of knots) Labastida-Marino-Ooguri-Vafa (LMOV) invariants. For fixed $j,\beta$ and $k$, the product over $l$ represents the quantum dilogarithm $(Q;q)_{\infty} = \prod_{l=0}^{\infty}(1-Qq^l)$.

The formulas (\ref{Z-closed-GV}) and (\ref{ZUV-prod}) are important examples of nontrivial string-theoretic conjectures. In the past two decades it has been verified in numerous examples that closed and open BPS multiplicities $N_{\beta,j}$ and $N_{\beta,k,j}$ encoded in these expressions are indeed integer. Until recently these checks were conducted only up to some order, and in consequence only for a finite set of BPS multiplicities. However, the  knots-quivers correspondence, discussed in the next section, enables to prove that certain infinite families of open BPS invariants are integer. Furthermore, let us stress that there is still no general mathematical definition of BPS numbers discussed above. Nonetheless, from a mathematical perspective, the expressions (\ref{Z-closed-GV}) and (\ref{ZUV-prod}) mean that closed and open Gromov-Witten invariants satisfy certain  integrality relations -- so in this sense a physical insight leads to deep mathematical statements, in line with the philosophy of physical mathematics.


\section{Knots-quivers correspondence} \label{sec-knots}

In this final section we present how various historical developments from section \ref{sec-ichep} and other ideas mentioned above merge together in a recently discovered \emph{knots-quivers correspondence} \cite{Kucharski:2017poe,Kucharski:2017ogk}. As its name indicates, it provides a relation between two branches of mathematics: knot theory and a quiver representation theory. However, its physical interpretation has broader consequences and characterizes a wide class of topological string theories and open BPS states in brane systems discussed above, whose multiplicities are encoded in generating functions of the form (\ref{ZUV-prod}) \cite{Panfil:2018faz}. Namely, it turns out that these BPS states (typically infinite number of them) are bound states of a finite number of certain elementary BPS states, and the way in which such bound states are formed is encoded in a symmetric quiver. Recall that a quiver is a diagram that consists of several nodes, connected by arrows. 
A quiver is symmetric if, for all pairs of nodes, the number of arrows from one node to another one is the same as the number of arrows in the opposite direction.

From the physical perspective, a quiver that encodes interactions of elementary BPS states arises in the effective description of the system in question, in terms of a supersymmetric quiver quantum mechanics. Moreover, the nodes of a quiver in various systems of interest also have a direct interpretation. In the case of knots, the nodes are in one-to-one correspondence with generators of HOMFLY-PT homology \cite{Kucharski:2017ogk}. From the viewpoint of topological strings, the nodes represent certain elementary discs, whose counting is captured by Gromov-Witten theory \cite{Panfil:2018faz,Ekholm:2018eee}. 

To proceed, let us provide a few quantitative results. Suppose we have a quiver with $m$ nodes and $C_{i,j}=C_{j,i}$ arrows between nodes $i$ and $j$. We refer to $C_{i,j}$ as elements of a ``quiver matrix'' $C$. To such a quiver one associates a generating series \cite{Kontsevich:2010px}
\begin{equation}
P_C(x_1,\ldots,x_m)=\sum_{d_1,\ldots,d_m} \frac{(-q^{1/2})^{\sum_{i,j=1}^m C_{i,j}d_id_j}}{(q;q)_{d_1}\cdots(q;q)_{d_m}} x_1^{d_1}\cdots x_m^{d_m},   \label{P-C}
\end{equation}
where $d_i$ are non-negative integers, and it depends on a motivic parameter $q$ and generating parameters $x_1,\ldots,x_m$. This series has the following product decomposition
\be
P_C(x_1,\ldots,x_m)=
\prod_{(d_1,\ldots,d_m)\neq 0} \prod_{j\in\mathbb{Z}} \prod_{l=1}^{\infty} \Big(1 -  \big( x_1^{d_1}\cdots x_m^{d_m} \big) q^{l+(j-1)/2} \Big)^{(-1)^{j+1}\Omega_{d_1,\ldots,d_m;j}}.   \label{PQx-Omega}
\ee
In quiver representation theory, this product decomposition encodes motivic Donaldson-Thomas invariants $\Omega_{d_1,\ldots,d_m;j}$ (in the exponent). Roughly, these invariants are Betti numbers of certain representations of a quiver in question, and it is proven that they are non-negative integers.

We can present now the relation between quivers and topological string amplitudes. Namely, as found in \cite{Kucharski:2017poe,Kucharski:2017ogk}, open topological string partition function given in (\ref{Zopen}) and (\ref{ZUV-prod}) can be expressed in the form of the quiver generating series (\ref{P-C}) with an appropriate identification of parameters $x_i$ (so that a dependence on a single parameter $x$ is introduced). In case of knots, represented by (in general complicated) branes in the resolved conifold, this is a generating function of colored HOMFLY-PT polynomials (\ref{Z-homfly}) that is identified with (\ref{P-C}). Moreover, one can also identify (\ref{P-C}) with brane partition functions in more general Calabi-Yau manifolds; for a simple class of Aganagic-Vafa branes in toric threefolds without compact 4-cycles such an identification is found in \cite{Panfil:2018faz}.

The above identification has deep consequences. First, it implies that the corresponding product expansion (\ref{ZUV-prod}) is identified with (\ref{PQx-Omega}). It follows that open BPS invariants $N_{\beta,k,j}$ are expressed as combinations of motivic Donaldson-Thomas invariants $\Omega_{d_1,\ldots,d_m;j}$. Since the latter invariants are proved to be integer, it must be so also for open BPS invariants, which proves a long-standing conjecture of their integrality in quite a general setting. 
Furthermore, in case of knots, from mathematical viewpoint, colored HOMFLY-PT polynomials are expressed in terms of numerical invariants of quiver moduli spaces, which can be interpreted as a novel type of categorification. Moreover, the identification of (\ref{Z-homfly}) with (\ref{P-C}) implies that the whole infinite family of colored HOMFLY-PT polynomials is determined by a finite number of parameters, i.e. the quiver matrix $C$ and parameters in the identification of $x_i$. For other consequences of the relation to quivers see \cite{Kucharski:2017poe,Kucharski:2017ogk,Panfil:2018faz,Ekholm:2018eee,Stosic:2020xwn}.

The knots-quivers correspondence has been verified for all knots up to 6 crossings, certain infinite families of torus knots and twist knots, rational knots, and more generally for a large infinite family of arborescent knots \cite{Stosic:2020xwn}. Providing its general proof for all knots, as well as generalization to yet more general branes and Calabi-Yau manifolds, are important challenges. Nonetheless, all manifestations of this correspondence revealed to date are already quite inspiring. As we hoped to convince the reader, they also nicely illustrate the ideas of physical and quantum mathematics.



\bigskip

\noindent
{\bf Acknowledgments} This work has been supported by the TEAM programme of the Foundation for Polish Science co-financed by the European Union under the European Regional Development Fund (POIR.04.04.00-00-5C55/17-00).


\end{document}